%% file: Taipei_Review_zhukovska.tex
\documentclass{article}

\usepackage[a4paper]{geometry}
\usepackage{txfonts,color, graphicx}
 \usepackage[pdftex,
        colorlinks=false,
        urlcolor=blue,       
        filecolor=green,     
        linkcolor=red,       
        pdftitle={Paper_dwarf},
        pdfauthor={Zhukovska},
        pdfsubject={article},
        pdfproducer={pdfLaTeX},
        pdfpagemode=None,
        bookmarksopen=true]{hyperref}

\begin{document}

\newcommand{\pos}[1]{\href{http://pos.sissa.it/cgi-bin/reader/contribution.cgi?id=#1}{\tt #1}}
\newcommand{\Zs}{\ensuremath{Z_{\odot}}}
\newcommand{\Ms}{\ensuremath{M_{\odot}}}
\newcommand{\Mspc}{\ensuremath{M_{\odot}\,\rm pc^{-2}}}
\newcommand{\Mspcgyr}{\ensuremath{M_{\odot}\,\rm pc^{-2}\,Gyr}}
\newcommand{\Msyr}{\ensuremath{M_{\odot}\,\rm yr^{-1}}}
\newcommand{\pyr}{\ensuremath{\rm yr^{-1}}}
\newcommand{\mum}{\ensuremath{\mu \rm m}}
\newcommand{\cmc}{\ensuremath{\rm cm^{-3}}}
\newcommand{\kms}{\ensuremath{\rm \,km\,s^{-1}}}
\newcommand{\ddt}[1]{{{\rm d}\, {#1} \over{\rm d}\,t}}
\input{journals.tex}
\title{Life Cycle of Dust in the Magellanic Clouds and the Milky Way}


\author{Svitlana Zhukovska and Thomas Henning}
\date{}
\maketitle

     \vspace*{-1cm}    
 \begin{center}
      Max Planck Institute for Astronomy\\
       K\"onigstuhl 17, 69117 Heidelberg, Germany\\
\end{center}

\abstract{To a great extent, our understanding of the life cycle of dust is based on the observational and theoretical studies of the Milky Way and the Magellanic Clouds, which will be the topic of this contribution. Over past years, a large volume of observations with unprecedented spatial resolution has been accumulated for the Milky Way. It permits investigations of different stages of the life cycle of dust, from its formation in stellar sources to destruction in star-forming regions and supernovae shocks. Observations of dust emission, extinction, polarisation of light, and interstellar element depletions in the solar neighbourhood provide the most accurate constraints for the reference dust models applied to study extragalactic systems. However, global spatial studies of the circumstellar and interstellar dust are complicated in the Milky Way disk because of high extinction, confusion along the line of sight and large uncertainties in distances. In contrast, the favourable location in the sky and the proximity of the Magellanic Clouds allow detailed multi-wavelength studies of the dust-forming stellar populations and the investigation of variations of the interstellar grain properties for the entire galaxies. They enable the first comparison between the global stardust production rates from theoretical calculations and those from observations, which confirm discrepancy between accumulated stardust mass and observed interstellar dust mass -- ``the missing dust-source'' problem. Modelling of the life cycle of dust in the Large Magellanic Cloud showed that dust growth by mantle accretion in the ISM, a major dust source in the Milky Way, can be responsible for the existing dust mass in the LMC. We will present comparison of the dust input from different sources to the dust budgets of the Milky Way and the Magellanic Clouds, which reveals how the role of these dust sources depends on metallicity.}

\section{Introduction}
Our understanding of the nature and origin of interstellar  dust is to great extent based on the observational and theoretical studies in the solar neighbourhood. It provides a testing ground for various dust models. Their development has been stimulated over the past decades by increasing power of observational facilities and advances in laboratory studies of cosmic dust analogues. Dust models constrained by high-resolution observations (e.g., IR and microwave dust emission, interstellar extinction, scattering, polarisation, and interstellar element depletions) in diffuse regions of the local Milky Way have been successfully applied to investigate the dust content in many extragalactic systems (e.g., \cite{2007ApJ...657..810D, 2004ApJS..152..211Z, 1990A&A...237..215D} and many others).
The properties of dust grains inferred from the diffuse regions are however not universal. It has been long recognised that the chemical composition, size distribution, and solid-state structure of dust grains vary substantially across galactic environments (molecular clouds, hot gas, stellar outflows, and young stellar objects) \cite{1995A&ARv...6..271D, Draine:2003p1007}.  These variations reflect evolutionary changes which grains experience during their life cycle in the ISM.

\begin{figure}
\centering
\includegraphics[width=0.9\textwidth]{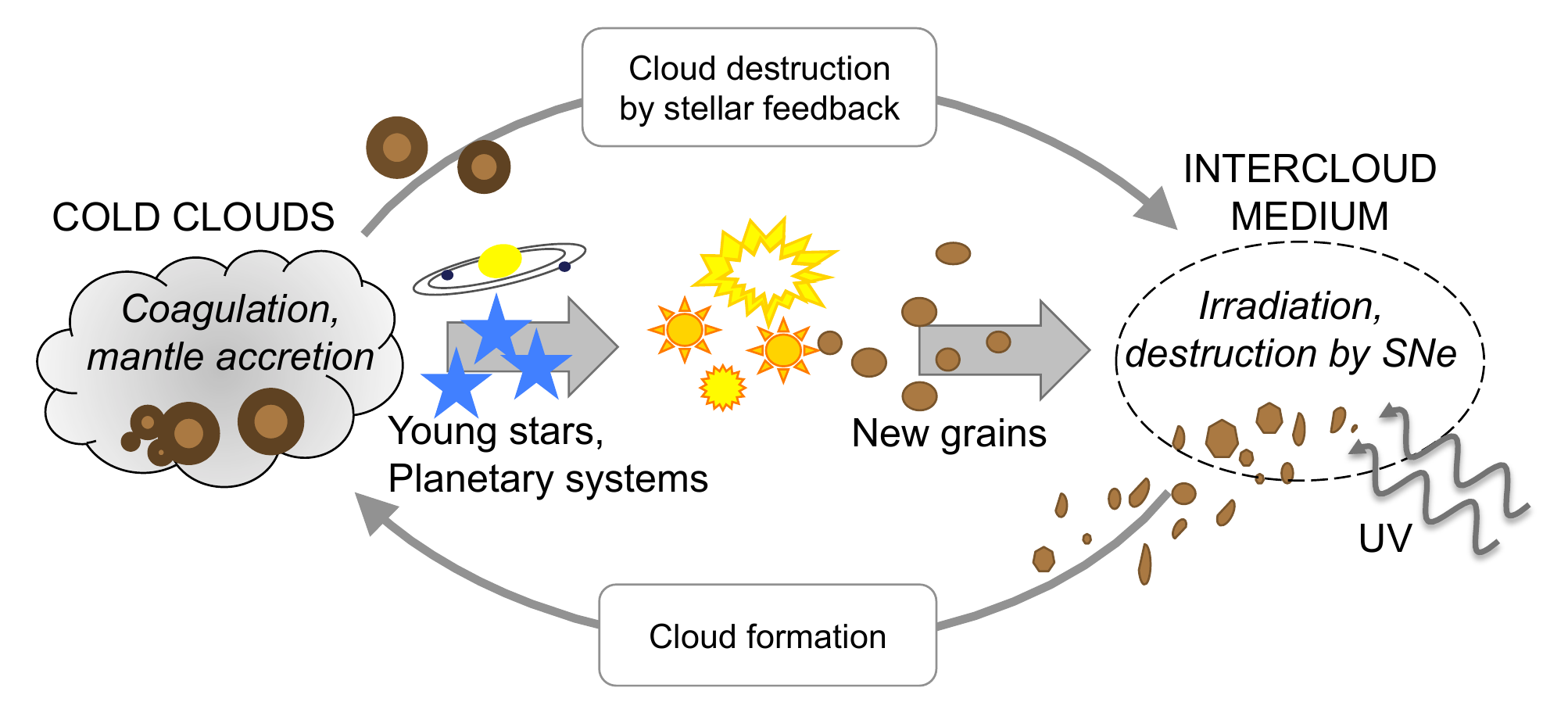}
\caption{Schematic representation of the gas and dust cycle between clouds, intercloud medium, and stars. The main processes affecting grains in the ISM are shown.}
\label{fig:lifecycle}
\end{figure}

A simplified scheme of the life cycle of interstellar dust  is shown in Fig.~\ref{fig:lifecycle}. Dust formation occurs at condensation temperatures in cooling outflows of evolved stars and expanding supernova ejecta through nucleation (formation of nm-sized molecule aggregates) and following grain growth to sub-\mum\ sizes. When stellar ejecta are mixed with the ambient ISM, newly condensed stardust grains become part of the interstellar dust population. In the diffuse phase, grains undergo various processing: ultraviolet (UV) irradiation, hydrogenation, and destructive processes resulting from grain-grain and gas-grain interactions in the interstellar shocks (melting, sputtering, vaporization, and shattering).
The interstellar dust takes part in the matter cycle between dense and diffuse clouds and intercloud medium governed by the processes of star formation and energetic stellar feedback.  The average lifetime of dust against destruction in the ISM (a few $10^8$~yr) is longer than the timescale of formation of molecular clouds and their disruption (a few $10^7$~yr). Grains therefore cycle multiple times between these phases before they are completely destroyed. In the dense clouds, grains are subject to grain growth by accretion of gas-phase species and coagulation, the processes which are responsible for large observed  variations in the shape of the extinction curve \cite{Fitzpatrick:2007p6352, HirashitaPoS}. 
Various dust processing across the ISM phase is reflected in the increase of interstellar element depletions with the gas density and decrease in the shocked gas observed in different sightlines \cite{Jenkins:2009p2144, Savage:1996p486, JenkinksPoS}. 
There is a direct evidence that a fraction of stardust grains ends their interstellar journey being incorporated in material forming new stars and protoplanetary disks. Stardust grains from the molecular cloud formed our solar system which survived its formation can be found as presolar grains in meteorites and analysed in laboratory (e.g., \cite{Clayton:2004dp, MessengerPoS}). Anomalous isotopic element ratios in these grains bear signatures of the nucleosynthesis of their parent stars permitting identification of their origin. The ratio of presolar grains of different origins derived in the laboratory can be used as a constraint on efficiency of dust production by stellar sources \cite{Zhukovska:2008bw}.

Despite the wealth of high-resolution observations in the local galactic environment, there is a number of obstacles in exploration of the entire life cycle of dust in the Milky Way. Our location in the galactic disk, high extinction in the disk, uncertainties in distances and  confusion along the line-of-sight hamper global studies of interstellar and circumstellar dust. These complications can be resolved for our neighbouring dwarf galaxies, the Large Magellanic Cloud (LMC) and the Small Magellanic Cloud (SMC). Their structure, kinematics, composition, distances and history have been studied observationally in great detail by ground-based and space observatories  over past decades \cite{Westerlund:1997va}. Moreover, their dust-forming stellar populations and the interstellar dust emission have been mapped from 3.6 to 500~\mum\ for the entire galaxies with Spitzer Space Science Telescope (SAGE survey) and Herschel Magellanic Survey (HERITAGE) \cite{2006AJ....132.2268M, Gordon:2011jq, Meixner:2013kr}. 
Another advantage is that the distances to the LMC (50~kpc) and the SMC (61~kpc) have been determined with high accuracy down to $\sim 3\%$ \cite{Pietrzynski:2013ck, Graczyk:2013is}. They  are located far away enough that the same distances can be assumed for all stars and interstellar gas clouds for intrinsic brightness determinations. 
 In addition, the metallicities of the LMC (0.3--0.5\Zs) and the SMC (0.2\Zs) \cite{Russell:1992eu} permit investigation of the lifecycle of dust at subsolar metallicities. Comparison of circumstellar dust in the SMC, the LMC and Milky Way gives a valuable insight into the metallicity dependence of dust formation process \cite{Gordon:2011jq, 2006AJ....132.2268M}. These metallicities are close to the metallicities in galaxies at the epoch of peak star formation at redshift of $\sim 1.5$. With their enhanced star formation rates and high gas fractions \cite{2004AJ....127.1531H, Harris:2009p7602}, the Clouds represent a template into the conditions for grain evolution in star forming galaxies at high redshifts \cite{Madau:1996tu, Tollerud:2011jn}.  
 The subsolar metallicity in the Clouds implies a lower interstellar dust content compared to the Milky Way, which in turn affects the physical conditions in the ISM. The mean free path of UV photons from massive stars is longer and intensity of the interstellar radiation field is therefore higher than in the local Milky Way. The molecular gas fractions in the LMC ($\sim 0.1$) and SMC ($\sim 0.05-0.10$) are therefore several times lower than that in the Milky Way (\cite{Leroy:2007da, Meixner:2013kr} and references therein). 

Some aspects of the lifecycle of interstellar dust in the Milky Way and Magellanic Clouds are covered in other contributions of this volume, to which we will refer where needed. This review will briefly describe the model of the lifecycle of dust and implementation of the main destruction and production processes and their relevant timescales in Sect.~\ref{sec:ModDustLife}. Section~\ref{sec:StellarSources} will focus on the dust input from AGB stars and type II SNe. We will discuss constraints on the stellar dust yeilds provided by the local Milky Way and metallicity dependence of the dust mixture from AGB stars by the Galactic disk. In the context of the global contribution of stars to the dust budget, theoretical and observational efforts to determine the dust production rates by AGB stars in the Magellanic Clouds will presented. Finally, the total contribution of stars to the interstellar dust budgets in the Milky Way and the Magellanic Clouds will be summarised in Sect.~\ref{sec:Final}.

\section{Models of the life cycle of dust in the ISM}\label{sec:ModDustLife}
Dwek and Scalo \cite{Dwek:1980p490} proposed a model of the life cycle of refractory grains in the solar neighbourhood that captures main dust formation and destruction processes operating in the ISM  (Fig.~\ref{fig:lifecycle}). This model is based on the chemical evolution model of the Galaxy, therefore it also accounts for the dependence of dust content on global galactic evolution processes such as star formation, and galactic in- and outflows. Other advantages of this approach is that it naturally includes element abundance constraints for the interstellar dust mixture, and allows to study variations of its amount and chemical composition. These quantities can be estimated from observations and compared with model predictions for the present day.  Since the original work \cite{Dwek:1980p490}, the models of the lifecycle of grains in the ISM have been further developed to include multiphase structure of the ISM \cite{Hirashita:2000p6583, Zhukovska:2008bw}, individual evolution of stardust grains of different origins \cite{Zhukovska:2008bw, Gail:2009p512}, variations in grain size distribution (see \cite{HirashitaPoS} in this volume) and applied for the Milky Way disk and galaxies of different morphological types in the local Universe \cite{Dwek:1998p67, Calura:2008p1752, Zhukovska:2013vg} and high redshifts \cite{Valiante:2011hu, Pipino:2011p6991, Gall:2011db}. Majority of dust evolution models include the evolution of two main dust constituents: silicate and carbonaceous grains. This dust mixture is built out of the most abundant refractory elements (``big five'': Mg, Si, Fe, O, and C) and satisfies extinction, emission and abundance constraints in the solar neighbourhood (e.g., \cite{2004ApJS..152..211Z}).

\subsection{Stellar dust production}
The global production rate of stardust of kind $i$ is determined from the dust and gas chemical evolution models by integration over stellar mass range ending their life at instant $t$ 
\begin{equation}
	\dot{M}_{i, \rm d}(t) =  
	 m_{\rm av}^{-1} \int_{m_l}^{m_u}{\phi(m) \psi  \left(t-\tau(m,Z)\right) Y_{i, \rm d}(m,Z)dm},
	\label{eq:DustInjRate} 
\end{equation}
where  $\phi(m)$ is the initial stellar mass function (IMF), $m_{\rm av} = \int m \phi(m) dm$ is the average stellar mass, $\tau(m,Z)$ is the stellar lifetime,  $\psi(t-\tau(m,Z)$ is the star formation rate at the instant of birth of a star with the initial mass $m$ and metallicity $Z$ in units, here $Z$ is the metallicity of the ISM  at the instant of birth of a star determined by galactic chemical evolution. $Y_{i,\rm d}(m,Z)$ is the dust yield,  i.e. the total mass of dust of kind $i$ returned to the ISM by a star  over its evolution. Mass- and metallicity-dependent dust yields are the central quantities for calculations of the total dust input from stars. The local Milky Way and the Magellanic Clouds provide complementary tests for the dust yields from both AGB stars and type II SNe as discussed in Sect.~\ref{sec:StellarSources}.  

\subsection{Dust evolution in the ISM}
The change of mass of dust species of type $i$ per unit time is 
\begin{equation}
	\ddt{M_{i, \rm d}} = \dot{M}_{i, \rm d}(t) + G_i - \frac{M_{i, \rm d}}{\tau_{i,\rm SN}} -  \frac{M_{i, \rm d}}{\tau_{i,\rm SF}}  + \dot{M}_{i, \rm d}^{\rm inf} - \dot{M}_{i, \rm d}^{\rm out},
\end{equation}
where the terms on the right side of the equation are the dust injection rate from stars given by Eq.~(\ref{eq:DustInjRate}), the dust growth rate by accretion in the ISM, the dust destruction rates by SNe and star formation, and two last terms denote an increase of the dust mass due to infall of matter from outside and decrease due to galactic outflows from the galaxy.

\begin{figure}
\centering
\includegraphics[width=0.75\textwidth]{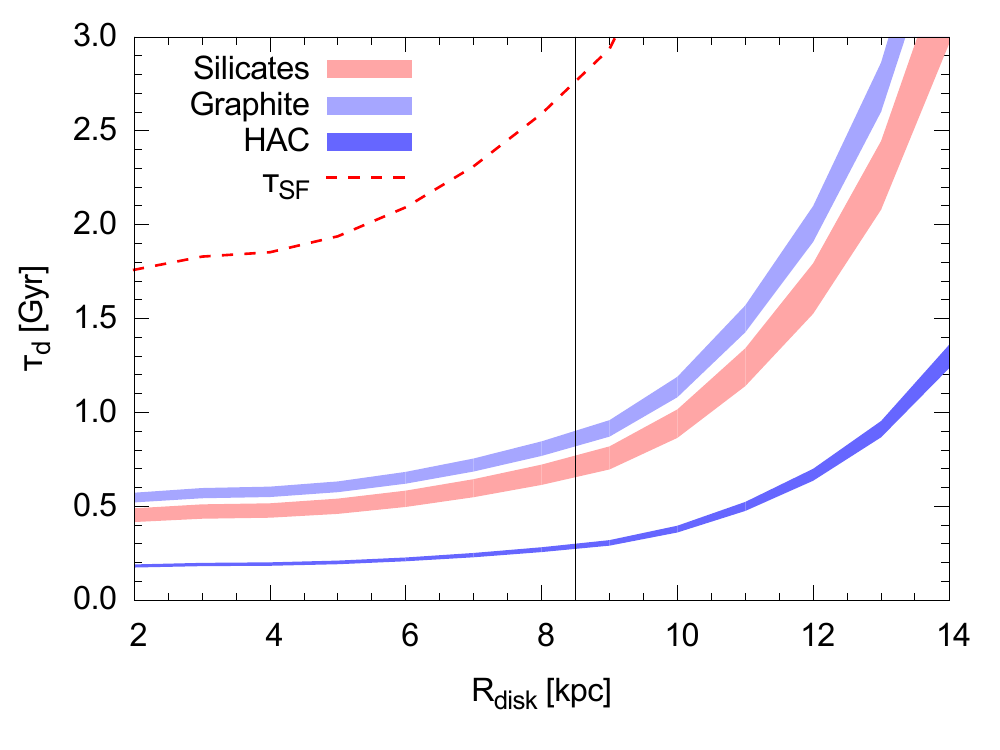}
\caption{Radial variations of the lifetimes of silicate (light red shaded area) and carbonaceous (graphite and HAC, light and dark blue areas, respectively) grains against destruction by SNe and star formation (dashed red line) in the present Milky Way disk. The vertical line marks the position of the solar system. The data are taken from \cite{Zhukovska:2008p7215}.}
\label{fig:taudest}
\end{figure}

\subsubsection{Dust destruction in the ISM}\label{sec:Destruction}
Dust grains are efficiently destroyed by SN blast waves predominantly in the warm neutral/ionised phase of the interstellar medium through inertial sputtering for shock velocities in the range of  50-150 km/s and through thermal sputtering at higher shock velocities \cite{Jones:1994p1037, Jones:1996p6593}. 
The global  lifetime of dust grains against destruction by SN shocks $\tau_{\rm SN} $ can be expressed in terms of the amount of interstellar gas cleared of dust by a single SN $m_{\rm cl}$ \cite{McKee:1989p1030}:
\begin{equation}
\label{eq:DuDestr}
\tau_{\rm SN} = \frac{M_{\rm gas}}{ m_{\rm cl} f_{SN} R_{SN}}\,,
\end{equation} 
where $M_{\rm gas}$ is the mass of gas in the galaxy, $R_{SN}$ is the net SN rate, and $f_{\rm SN}$ is the fraction of all SNe that destroy dust in ISM.  $f_{\rm SN}$ accounts for massive stars which end their life above the galactic gaseous disk and do not interact with dust. Additionally, the total SN rate is reduced due to correlated SNe exploding in existing superbubbles created by previous SNe in the cluster. The expansion velocity of the supershell created by the multiple SNe of the cluster is below the limit for dust destruction. At least $75\%$ of all supernovae  in the Galaxy are expected to occur in a correlated fashion \cite{Higdon:2005p1035}.  It is therefore necessary to account for the reduction of the total SN rate. $f_{SN}$ estimated from observations has similar values for the Milky Way and the LMC, 0.35 and 0.3, respectively \cite{McKee:1989p1030, Zhukovska:2013vg}. 
$m_{\rm cl}$ is calculated for a given SN shock structure and it depends on the properties of dust material, grain size distribution and density of the ambient ISM \cite{Jones:1994p1037}. The $m_{\rm cl} $ values of 1300~\Ms\ and 1600~\Ms\ for graphite and silicates, respectively, derived for the solar neighbourhood, are similar to the estimates in the LMC \cite{Zhukovska:2013vg}. 

For the Milky Way, Equation~(\ref{eq:DuDestr}) yields the average destruction timescales of $4\times 10^8$~yr  and  $6\times 10^8$~yr, for silicate and graphitic dust, respectively \cite{Jones:1994p1037}. If carbonaceous grains are assumed to be in form of hydrogenated amorphous carbon (HAC), their lifetime is shortened by a factor of three \cite{SerraDiazCano:2008p588}. The lifetime of the PAH molecules, another interstellar dust component, is even shorter, $1.4-1.6 \times 10^8$~yr \cite{Micelotta:2010dl}. These lifetimes are relatively short compared to the timescale of galactic chemical evolution. Efficient dust destruction by SN blast waves is supported by Spitzer Space telescope observations of interaction of the Puppis A supernova remnant in our galaxy with a neighbouring molecular cloud \cite{Arendt:2010p7737}. Modeling the spectra from regions in front of and behind the SN remnant shock front reveals that the ubiquitous polycyclic aromatic hydrocarbons (PAHs)  are completely destroyed in the swept-up ISM, along with nearly 25\% of the mass of graphite and silicate dust grains on the timescale of 80~yr.
Dust lifetimes are not the same throughout the Galactic disk, since chemical evolution of the disk is significantly slower in the outer regions than in the inner galaxy, as indicated by the at least 10 times lower star formation rate in the outer disk and large radial metallicity gradient. Figure~\ref{fig:taudest} shows the radial variations of the lifetimes for the main dust components, silicate and carbonaceous dust, in the Galactic disk calculated using the gas mass and SN rate per unit surface area from the galactic chemical evolution model in Eq.~(\ref{eq:DuDestr}). The life cycle of grains is more than 6 times faster in the inner Galaxy compared to the outer disk as seen in Fig.~\ref{fig:taudest} \cite{Zhukovska:2008p7215}.

The grain lifetimes derived from Eq.~(\ref{eq:DuDestr}) for the LMC are equal to 0.8~Gyr for silicate and 1~Gyr for graphitic grains \cite{Zhukovska:2013vg}. This is somewhat longer than for the solar neighbourhood, see Fig.~\ref{fig:taudest} for comparison. Using $M_{\rm g}=5.7\times10^8$~yr and $R_{\rm SN}=2.9 \times 10^{-3}\ \rm yr^{-1}$ for the SMC and adopting for $m_{\rm cl}$ and $f_{\rm SN}$ the same values as for the LMC \cite{Zhukovska:2013vg}, one yields values of the dust lifetimes of 0.4 and 0.5~Gyr, for silicate and graphitic dust, respectively, which are close to those in the Milky Way. Note that there are large uncertainties in estimates of the global galactic lifetimes from Eq.~(\ref{eq:DuDestr}) related to uncertainties in the observational estimates of gas masses and SN rates \cite{2011A&A...530A..44J}.
Additionally, the local values of the lifetime of grains against destruction can deviate from the average value, in particular, in regions that actively form stars. Star formation histories of the LMC and the SMC reveal that such regions are nonuniformly distributed both temporarily and morphologically in these galaxies \cite{2004AJ....127.1531H, Harris:2009p7602}.  
On the other hand, stellar element abundances indicate a low if any metallicity gradient in both the LMC and the SMC, with exception of the oldest stellar populations \cite{Cioni:2009db, Feast:2010p7238}. This implies an efficient turbulent mixing of metals in the ISM which may mitigate inhomogeneous distribution of dust due to the enhanced destruction in star forming regions. 

Interstellar dust is also destroyed in the process of star formation on the timescale $\tau_{\rm SF}$ determined by the ratio of the gas mass to the star formation rate. Its variation with the Milky Way radius are also shown in Fig.~\ref{fig:taudest}.  The value of $\tau_{\rm SF} \sim 2.5$~Gyr estimated for the solar neighborhood is somewhat lower than 3.1~Gyr estimated for the LMC  \cite{Zhukovska:2013vg}. The SMC has higher gas fraction and lower star formation rate than the LMC and Milky Way, therefore $\tau_{\rm SF}\gtrsim 10$~Gyr in the SMC is longer than in these galaxies. 

\subsubsection{Dust growth in the ISM}
Many works have found that additional dust growth by accretion of gas-phase species on existing grain surfaces in the ISM is required to explain high levels of element depletions in the ISM (e.g., \cite{Dwek:1980p490, Dwek:1998p67, Zhukovska:2008bw, Draine:2009p6616, JenkinksPoS} and references therein). Nevertheless, the details of these process are not well-understood. Observational evidence of amorphous silicate and carbonaceous dust reside as two distinct populations seems to imply that the growth process should be selective: Si, Fe, Mg and O atoms should preferentially stick to the silicate grain surfaces and C atoms to the carbonaceous dust surfaces, in contrast to what is expected for low-temperature condensates. 
\cite{Draine:2009p6616} suggested that selective growth could occur in the diffuse ISM in the presence of UV photons. In this scenario, the UV irradiation helps to rearrange chemical bonds to form amorphous silicates, as Si, Mg, Fe, O atoms are adsorbed on the silicate grain surface, and remove C atoms by the UV photoexcitation. Similarly, carbon dust could grow by addition of impinging C atoms, while other species arriving on the carbon dust surface are removed by the UV photons \cite{Draine:2009p6616}. Recently, a new  detailed model of the life cycle of carbonaceous grains have been proposed, in which carbon dust in the form of hydrogenated amorphous carbon grows (re-forms) by accretion in the diffuse outer regions of dense molecular clouds (see \cite{Jones:2014in} and \cite{JonesPoS} for recent reviews). For silicate grains, the growth mechanism is less clear. Recent experiments on low-temperature condensation has demonstrated that the formation of solid SiO is possible at a low temperature without energy barrier \cite{Krasnokutski:2014bi}. Since SiO is the main building block of astronomical silicates, this finding  is an essential step forward to understanding of the formation of more complex silicates.

Dust evolution models usually assume that the dust growth occurs by collisions with gas-phase species in dense clouds on the existing grains. The corresponding equation for the change of condensation degree of a dust-forming species of kind $i$ in a dense cloud is
\begin{equation}
	\ddt{f_i} = {f_i (1-f_i)\over \tau_{i, \rm acc}}\,,
	\label{eq:CondDegree}
\end{equation}
where $\tau_{i, \rm acc} $ is the the accretion timescale. It is the key quantity for dust growth, which depends on the properties of the grain solid material, grain size distribution, and the physical conditions in the gas:
\begin{equation}
           \tau^{-1}_{i,\rm acc} = {3 \alpha_i A_{i} m_{\rm amu}\over \rho_{i,\rm c} \nu_{i,\rm c}} \cdot {1\over \langle a_i \rangle} \cdot \upsilon _{i,\rm th} n_{\rm H}   \cdot \epsilon_i \, ,
\label{eq:TauGrowthDu}
\end{equation}
where 
$\alpha_i$ is the sticking efficiency to the grain surface, 
$A_{i}$ is the atomic weight of one formula unit of the dust species,
$\rho_{i,\rm c} $ is the density and $\nu_{i,\rm c} $ is the number of atoms of the key element contained in the formula unit of the condensed phase,
$\langle a_i\rangle$ is the average grain radius determined by the grain size distribution,
$\upsilon _{i,\rm th}$ is the thermal speed of the growth species, 
$n_{\rm H}$ is the number density of H nuclei,
and $\epsilon_i$ is the element abundance of the key species.

\begin{figure}[t]
\centering
\includegraphics[width=0.49\textwidth]{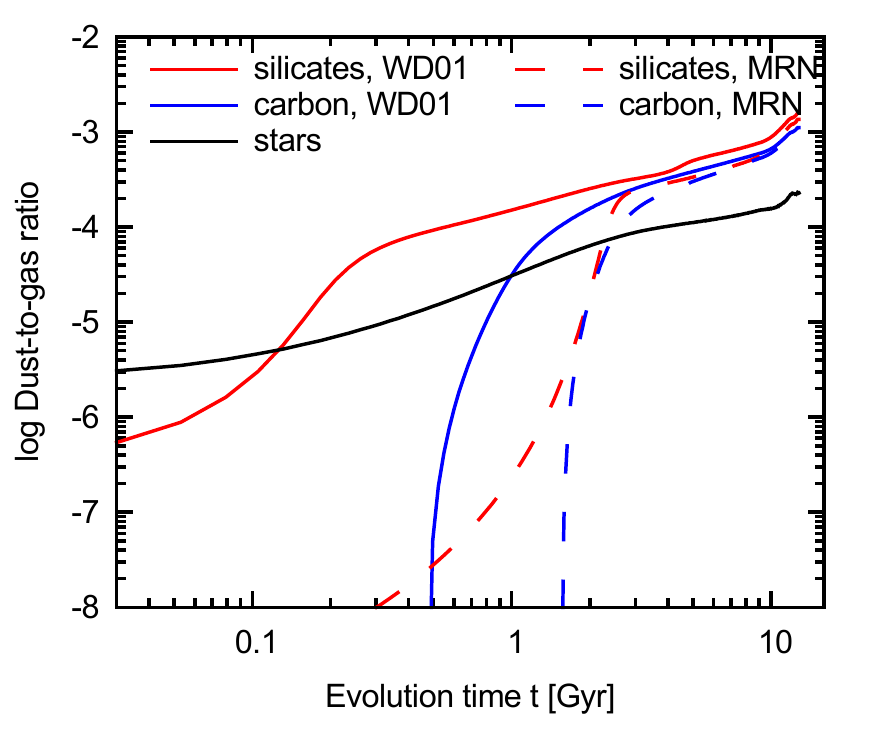}
\includegraphics[width=0.49\textwidth]{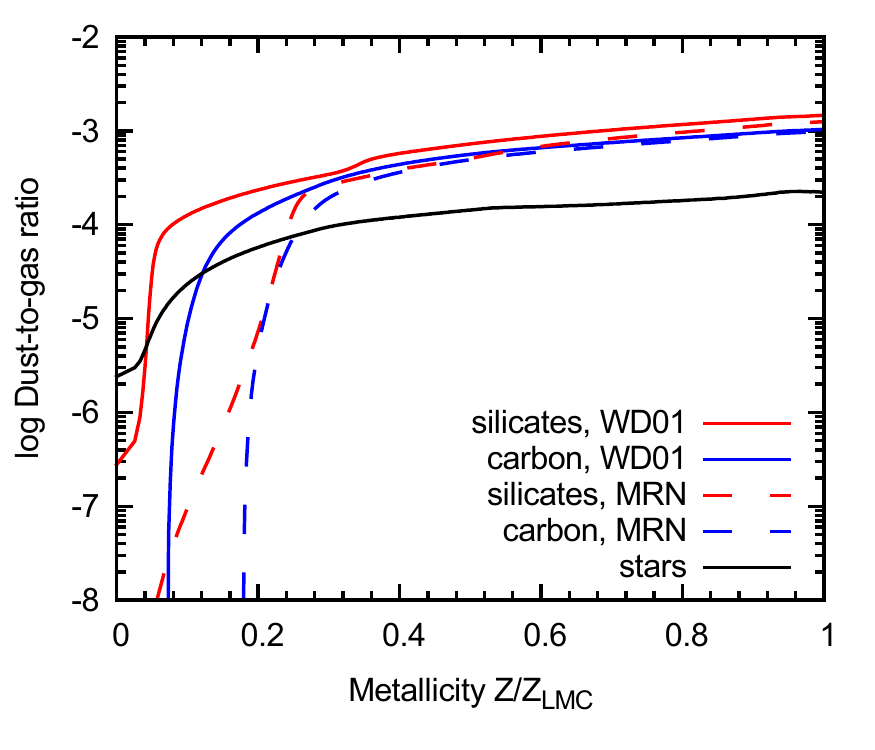}
\caption{Evolution of the dust-to-gas ratio in the LMC calculated for two grain size distribution: a simple power law from \cite{Mathis:1977p750} (MRN, dashed lines) and a more complex distribution from \cite{Weingartner:2001p475} derived from dust extinction studies (WD01, solid lines), as a function of time and metallicity (left and right panels, respectively). The dust-to-gas ratio is shown separately for the interstellar silicate and graphitic grains (red and blue lines, respectively) and for the stardust, which includes dust from AGB stars and SNe (black line).}
\label{fig:DGR}
\end{figure}
 

The dust growth timescale is inversely proportional to the abundance of the key species. It is therefore very inefficient in metal-poor environment until the metallicity in the ISM reaches some critical value \cite{Zhukovska:2008p7215, Zhukovska:2008bw, Zhukovska:2009p7232, Asano:2013kl}. In the local Milky Way, the critical metallicity for the dust growth is reached within a few 100~Myr evolution. In metal-poor dwarf galaxies, the growth is slower and it may take up to roughly 1~Gyr, for the dust growth to become important \cite{Zhukovska:2014ey}. The grain size distribution the Magellanic Clouds, characterised by a larger abundance of small grains (a larger grain surface area) compared to the Milky Way \cite{Weingartner:2001p475}, can somewhat mitigate the effect of their low metallicities on the growth timescale, accordingly to Eq.~\ref{eq:TauGrowthDu}. In order to illustrate the effect of the grain size distribution on the dust content of a galaxy, we perform calculations of the dust evolution in the LMC assuming two different size distributions: the standard power law commonly used for dust in the local diffuse ISM \cite{Mathis:1977p750} and a more complex size distribution derived from fitting of the average extinction curve in the LMC \cite{Weingartner:2001p475}. The resulting dust-to-gas ratio is shown separately for the silicate and carbon dust in Fig.~\ref{fig:DGR}. Time evolution of the dust-to-gas ratio is indeed sensitive to the assumed grain size distribution, but only during the first 1~Gyr corresponding to the metallicity of 0.003. When dust production by stars and by growth in the ISM reaches the balance with dust destruction in the ISM, the average dust abundances in the ISM do not depend on the grain size distribution.

Several approaches to calculation of the dust production rate have been suggested in the literature. 
One can directly use Eq.~(\ref{eq:CondDegree}) to calculate the dust production rate $G_i$ as suggested in  \cite{Dwek:1998p67}. Other approaches explicitly account that the dust growth occurs at high densities in molecular clouds, which are characterised by the  lifetime $\tau_{\rm cl}$ and the total mass fraction $X_{\rm cl}$(e.g., \cite{Hirashita:2011jr, Zhukovska:2008bw}). The dust production rate can be expressed as \cite{Zhukovska:2008bw}
\begin{equation}
G_{j,\rm d}={1\over\tau_{\rm exch,eff}}\Bigl[\,f_{j,\rm ret}M_{j,\rm d,max}-
M_{j,\rm d}\,\Bigr]\,,
\label{eq:DuProdInCloud}
\end{equation}
where $\tau_{\rm exch,eff} = (1-X_{\rm cl})/X_{\rm cl} \tau_{\rm cl}$ is the effective exchange time required to cycle all ISM through clouds, $f_{j,\rm ret}$ is the condensation degree for growth species of kind $j$ after cloud dispersal, $M_{j,\rm d,max}$ is the maximum possible dust mass assuming complete condensation, $M_{j,\rm d}$ is the current average dust mass.

\section{Dust production by stellar sources}\label{sec:StellarSources}
Low- and intermediate-mass stars ($m_* \lesssim 8\,\Ms$) at asymptotic giant branch (AGB) and massive stars ($m_* \gtrsim 8\,\Ms$) after the final SN explosion become the main stellar factories of dust, but their importance for the dust content of galaxies is a  long-standing debate. The local Milky Way and the Magellanic Clouds offer unique complementary constraints for the models of the life cycle of dust permitting to ascertain the net dust input from stellar sources to the galactic dust budget. The star formation histories and chemical evolution of these galactic systems were extensively studied in many observational works providing a reliable basis for the dust evolution models described in the preceding section. 
In the following we will summarise recent advances in theoretical and observational studies of contributions of stellar sources to the dust budget in the solar neighborhood, Milky Way disk and Magellanic Clouds.

\begin{figure}[t]
\centering
\includegraphics[width=0.75\textwidth]{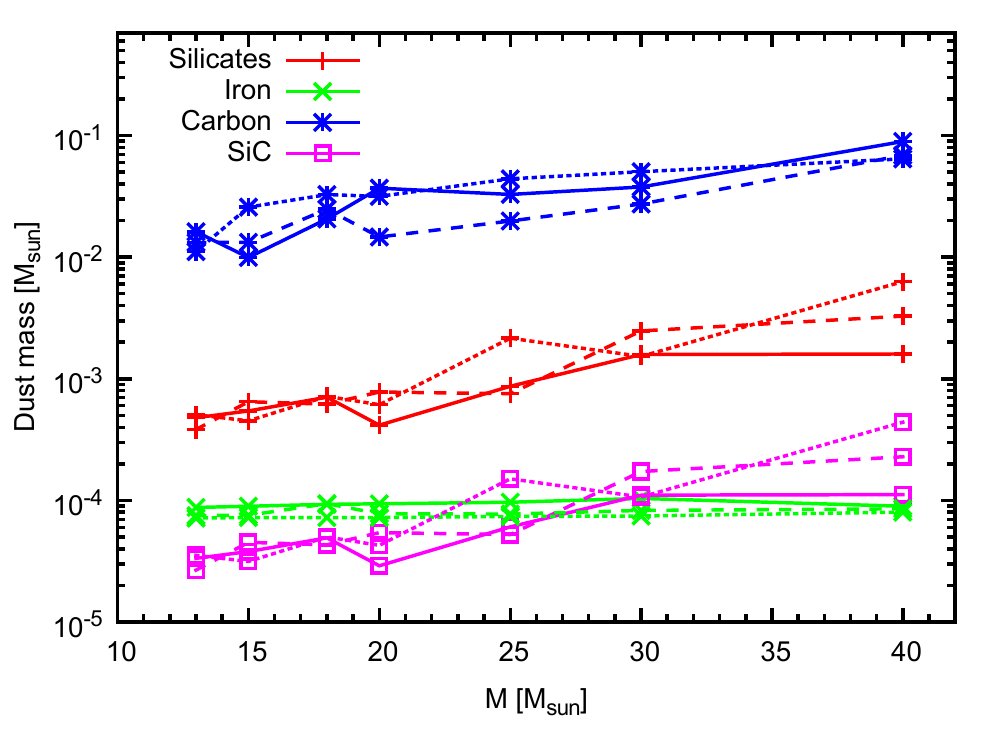}
\caption{Dust masses returned by type II SNe as a function of progenitor mass calculated with the fixed condensation efficiencies for silicate, SiC, and carbon dust derived in \cite{Zhukovska:2008p7215, Zhukovska:2008bw} with aid of dust evolution models. The solid, dashed and dotted lines correspond to the progenitor metallicities of 0.02, 0.004 and 0, respectively.}
\label{fig:SN}
\end{figure}

\subsection{Role of type II SNe in dust production} 
Hot debates about efficiencies\footnote{Efficiency or degree of condensation is defined as a ratio of the mass of refractory elements condensed into dust to their total mass in ejecta.} of dust condensation in ejecta of core-collapse SNe have been revived with recent discoveries of copious dust masses in objects in the early Universe. Given that type II SNe produce heavy elements on a short timescale, they are potentially the primary source of dust at high redshifts \cite{Gall:2011hr}. SN remnants in the Milky Way and Magellanic Clouds are subjects of many observational studies of dust formation in SNe stimulated by new windows into far-IR wavelength opened by Herschel Space Telescope and into sub-/millimitre by ALMA. 
Analysis of the IR dust emission in a few available data for SN remnants yields low values of condensation efficiencies in type II SNe of $10^{-4} - 10^{-3}$ (e.g., \cite{Rho:2008p4396, Rho:2009p4397, Temim:2012vj}, see \cite{GomezPoS} in this volume for a review). Recent far-IR and millimetre observations of historical SN remnants Cassiopeia A and Crab Nebulae revealed larger dust masses of 0.1-0.7~\Ms\ hidden in the cold component  \cite{Matsuura:2011ij, Gomez:2012fm, GomezPoS}. These discoveries do not however necessary imply that core-collapse SNe are the main dust factories in galaxies for two reasons.
First, it is unclear what fraction of these grains will survive destruction in reverse shocks of SN remnants. Numerical simulations indicate that a large fraction up to 100\% of freshly condensed dust can be destroyed by the reverse shock \cite{Silvia:2010p6576, Silvia:2012br, Bianchi:2007p2222}.  Size distribution of the SN condensates is crucial for grain survival, which increases with the grain size. Secondly, if one SN destroys dust in  1300--1600~\Ms\ of gas as discussed in Sect.~\ref{sec:Destruction}, it has to produce a similar dust amount,  about $13-16\Ms$ assuming a dust-to-gas ratio of 1\%, for the positive net input. This value by far exceeds the largest estimates from observations.

\begin{table}
\centering
\caption{Global production rates of C-rich and O-rich dust from various stellar sources in the solar neighbourhood in units $\Ms\,\rm kpc^{-2}Myr^{-1}$}
\smallskip
\label{tab:MW_DPR}
\begin{tabular}{l l l l}
\hline\hline
\noalign{\smallskip}
 Stellar source         & Dwek 1998 \cite{Dwek:1998p67}	& Tielens et al. 2005 \cite{Tielens:2005p1013} & Zhukovska et al \cite{Zhukovska:2008bw}\\
 \hline\noalign{\smallskip}
C-rich dust, AGB stars   &   2.8 &  3   &  1.0 \\
O-rich dust, AGB stars   &   3.7 &  5	&  0.4 \\
C-rich dust, SNe II	    &   1.5 & $<2$  & 0.7 \\
O-rich dust, SNe II 	&   7.0 & $<10$ & 0.01 \\
Type Ia SNe           &	 3.5	& $<2$ & 	0.002 \\
Wolf-Rayet  stars   & 	0.02	& $<0.06$ &	- \\
Novae	         & $<0.003$	& $<0.3$	& -\\
\hline
\end{tabular}
\end{table}

Dust formation in SNe can be studied not only by observations and modelling of SN remnants, but also by analysis of actual SN grains extracted from meteorites in laboratory. These stardust grains from type II SNe and AGB stars, or so-called presolar grains, are fossils of the material which formed the solar system 4.56~Gyr ago. Their origin is identified by anomalous isotopic element ratios bearing nucleosynthesis signatures of their parent stars. The meteoritic data yield an abundance ratio of dust from AGB stars and SNe at instant of the solar system formation, which can be used to constrain the condensation degree in SNe for available dust yields in AGB stars. The total rate of dust production can be calculated from Eq.~(\ref{eq:DustInjRate}) by integration over stellar masses below 8~\Ms\ for AGB stars and above 8~\Ms\ for SNe. Fitting the abundance ratios of stardust from AGB stars and SNe measured in laboratory in presolar grains with the ratio predicted by dust evolution models, one derives    the values of the condensation degrees required to reproduce this ratio \cite{Zhukovska:2008bw}. 
The fraction of refractory elements condensed in dust derived with such method is 0.15 for carbonaceous grains, $10^{-3}$ for silicates and $10^{-4}$ for SiC (\cite{Zhukovska:2008p7215, Zhukovska:2008bw}). These condensation fractions correspond to the dust masses shown in Fig.~\ref{fig:SN} as a function of progenitor mass for metallicities $Z=0,$ 0.004, and 0.02. For carbonaceous dust, the main dust species condensed in SNII, the condensed mass varies from $\sim 10^{-2}$ to 0.1\Ms, with a weak metallicity dependence.
The fact that these masses are lower than the masses of cold dust in SN remnants obtained by recent observations \cite{Matsuura:2011ij, Gomez:2012fm} may imply partial destruction of newly formed grains in the reverse shock. A relation between the dust-to-gas ratio and metallicity derived from observations of a large dwarf galaxy sample requires the low net input from core-collapse SNe.

Table~\ref{tab:MW_DPR} shows the net dust production rates from core-collapse SNe in the solar neighbourhood for the present epoch estimated in several works. A large discrepancy between model predictions from \cite{Zhukovska:2008bw} and from \cite{Dwek:1998p67} and \cite{Tielens:2005p1013}  is caused by the almost complete condensation of refractory elements in dust assumed in these works. The low condensation fraction of silicates inferred from meteoritic studies substantially reduces the total dust mass condensed in SN ejecta, so that AGB stars appears to be the dominant the stellar source of dust.

Various estimates of dust production rates (DPR) from SNe in the Magellanic Clouds are given in Table~\ref{tab:MCs_DPR}. Models of dust evolution with the conservative condensation efficiencies from \cite{Zhukovska:2008p7215} indicate that dust input in the LMC is also dominated by AGB stars. The DPR for carbonaceous and silicate dust are $4.5\times 10^{-5}\Msyr$ and $0.2\times 10^{-5}\Msyr$, respectively. Recent estimates based on the SN rates and the condensed dust mass per SN determined by observations yield values of $(0.7-400)\times10^{-5}$ for the LMC and $(0.1-110)\times 10^{-5}\Msyr$ for the SMC \cite{Boyer:2012ck, 2013MNRAS.429.2527M}. Large uncertainties in these estimates are caused by existing controversies in the dust masses derived from observations, which vary from $10^{-4}\Ms$ to 0.5\Ms\ \cite{Matsuura:2011ij}.

\subsection{Contribution of AGB stars to the interstellar dust budget}
 Theoretical estimates of the net dust input from low- and intermediate-mass stars strongly depend on the adopted dust yields \cite{Zhukovska:2013vg}. Mass- and metallicity-dependent dust yields for AGB stars are derived from the state-of-the-art numerical calculations which combine extensive models for condensation of the multicomponent dust mixture in stellar winds with stellar evolution models, for the first time, performed by \cite{Ferrarotti:2006p993}. Stellar evolution prescription for the central star is essential for the amount and composition of the condensed dust as it determines the variations of the chemical surface composition during stellar evolution along the AGB and thus the chemistry of the wind. Recently, a number of new model calculations have been published in the literature, which differ in the approaches to stellar evolution modelling (for details see \cite{Nanni:2013wb, Ventura:2012cs} and references therein). In the following we will discuss the stardust injection rates from AGB stars in the solar neighbourhood and their comparison with presolar grain studies, variations of stellar dust production in the Galactic disk, and the net dust input from stars in the Magellanic Clouds, from observational and theoretical perspective.

\begin{table}
\caption{Global production rates of C-rich and O-rich dust from AGB stars and type II SNe in the Magellanic Clouds in units $\Ms\,\rm yr^{-1}$ derived by the dust evolution model in the LMC \cite{Zhukovska:2013vg} and IR observations within SAGE-LMC and SAGE-SMC programs \cite{Srinivasan:6p7509, Matsuura:2009p363, Boyer:2012ck, Riebel:2012eq, 2013MNRAS.429.2527M}.}
\smallskip
\label{tab:MCs_DPR}
\begin{tabular}{l l l l }
\hline\hline
\noalign{\smallskip}
 Type of dust, source         &  LMC   & LMC & SMC \\
                                         &  \textit{model}   & \textit{observations} & \textit{observations} \\
 \hline\noalign{\smallskip}
AGB stars, C-rich dust   & $5.7\times 10^{-5}$ &  $ (0.95 - 4.3) \times 10^{-5}$ &  $(0.8 - 4) \times 10^{-6}$  \\
 AGB stars, O-rich dust   & $1.3\times 10^{-6}$ &  $(0.95^a - 5.5)\times 10^{-6}$   & $(0.08 - 3) \times 10^{-6}$  \\
Type II supernovae	    & $4.7\times 10^{-5\,b}$   & $(0.7-400)\times 10^{-5\,c}$   & $(0.1-110)\times 10^{-5\,c}$ \\
\hline
\end{tabular}

\medskip
\begin{minipage}{0.95\textwidth}
{\scriptsize
$^a$ For the lower limit,  we added the dust production rates from anomalous and regular O-rich AGB stars from \cite{Boyer:2012ck}.\\
$^b$The value includes the carbon dust injection rate of $4.5\times 10^{-5}\Msyr$ and silicate dust mass of $2.0\times 10^{-6}\Msyr$ \cite{Zhukovska:2013vg}.\\
$^c$ The range in SN DPR is computed assuming that SN input $10^{-3}-0.5 \Ms$ of dust, taking the range of values from \cite{Boyer:2012ck, 2013MNRAS.429.2527M}. The mass of dust detected in SN 1987A is taken for the upper limit assuming the dust will not be destroyed by the reverse shock.
}
\end{minipage}
\end{table}

\begin{figure}[t]
\centering
\includegraphics[width=0.75\textwidth]{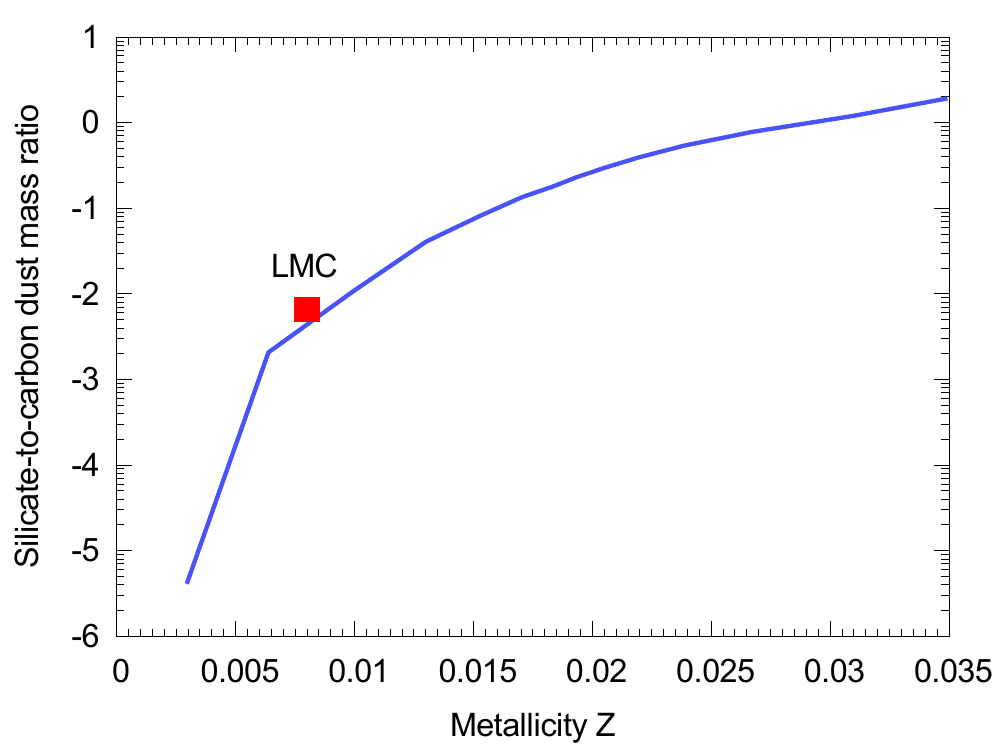}
\caption{Silicate-to-carbon dust mass ratio in interstellar grain population from AGB stars as a function of present-day metallicity in the Galactic disk (solid line) and LMC (red square). The data for the figure are adopted from the dust evolution models for the Milky Way and for the LMC  \cite{Zhukovska:2008p7215, Zhukovska:2013vg}.}
\label{fig:AGBradial}
\end{figure}

\subsubsection{Solar neighbourhood}
The rates at which AGB stars enrich the local ISM with new C- and O-rich dust estimated by several authors \cite{Dwek:1998p67,Tielens:2005p1013, Zhukovska:2008bw} are listed in Table~\ref{tab:MW_DPR}. The main reason for the large differences between the values in Table~\ref{tab:MW_DPR} is \cite{Dwek:1998p67, Tielens:2005p1013} assumed the almost complete condensation of refractory elements in AGB ejecta, while \cite{Zhukovska:2008bw} employed the dust yields based on theoretical models \cite{Ferrarotti:2006p993}, which predict incomplete condensation. Accordingly to the  model calculations from \cite{Zhukovska:2008bw}, the dust input from AGB stars is dominated by carbon-rich dust, which is injected into the ISM at the rate of $1\ \Ms \rm kpc^{-2} Myr^{-1}$. This is 2.5 times higher than the production rate for silicate dust \cite{Zhukovska:2008bw}. 

Over the years, the efforts of the meteoritic science community on analysis of anomalous isotopic ratios in presolar dust grains  has accumulated a wealth of data on the masses and metallicities of their parent stars. With the models of dust evolution combined with $m$- and $Z$-dependent yields one can determine the stellar population of AGB stars that contribute to the stardust inventory at the instant of the solar system formation. Comparison of the mass and metallicity distribution of parent AGB stars inferred by presolar grain studies with those predicted by the models provides a valuable test for the evolutionary dust models \cite{Gail:2009p512}. Good agreement between the predictions from the dust model and the findings from the isotopic compositions of presolar grains is found for SiC grains. The majority of presolar SiC grains originates from AGB stars with masses from about 1.5 to 3--4\Ms\ and of roughly solar metallicity. Only a small fraction of SiC grains are from stars with metallicity lower than about 0.5\Zs. The predicted mass distribution of AGB stars that contributed O-rich dust is essentially bimodal, with roughly equal contributions from stars in the ranges 1.3--2.5 and 4--8~\Ms\ of about solar or slightly subsolar metallicities. The meteoritic data indicate that a large fraction of O-rich dust indeed comes from AGB stars with masses from about 1 to 2.5~\Ms. However, presolar grains with isotopic anomalies of hot-bottom burning characterising massive AGB stars are completely missing from the presolar grain inventory \cite{Gail:2009p512}. This puzzling discrepancy remains to be understood.

\subsubsection{Milky Way disk}
Presence of the galactic radial metallicity gradient allows to investigate how the population of AGB stars and their dust production depend on metallicity.  The radial distribution of evolved stars derived from IR data from IRAS and later from AKARI space missions has a flat profile for carbon-rich stars and exponential distribution for  oxygen-rich stars \cite{Ishihara:2011ip, Thronson:1987p4295}. This is in a good agreement with the radial distributions of C- and O-rich dust from AGB stars predicted by the model of the life cycle of grains in the Milky Way \cite{Zhukovska:2008p7215}. 

The chemical composition of the stardust population from AGB stars residing in the ISM is significantly more C-rich compared to the interstellar dust. The radial profile of the current mass ratio of silicate to carbonaceous interstellar grains of AGB origin is shown as a function of the ISM metallicity in the  disk in Fig.~\ref{fig:AGBradial}. For comparison, the silicate-to-carbon dust mass ratio in the solar neighbourhood is about 1.7. The dust mixture from AGB stars approaches this values only in the inner Galaxy at galactocentric disk radii $R<4$~kpc and $Z\gtrsim 2\Zs$. In most of the Galaxy, dust population from AGB stars remain C-rich. The silicate-to-carbon ratio depends non-linearly on metallicity and its slope decreases with $Z$. A steeper slope at lower metallicities in the outer disk partly results from the strong non-linear metallicity dependence of dust condensation in stellar outflows. Additionally, the lifetimes of dust against destruction is longer in the outer disk permitting  accumulation of grain from many generations of more metal-poor stars, which produce more carbon-rich dust.

\begin{figure}
\centering
\includegraphics[width=0.49\textwidth]{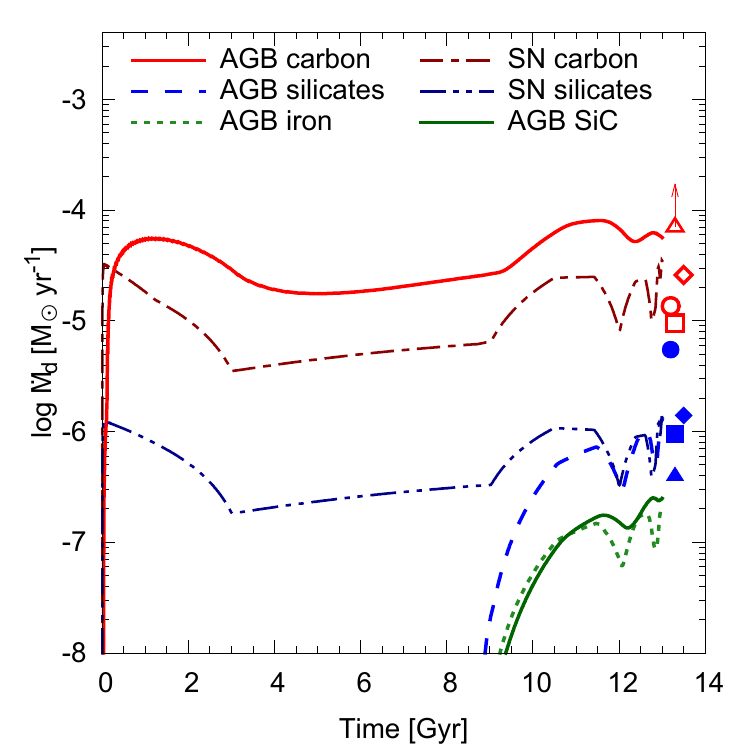}
\includegraphics[width=0.49\textwidth]{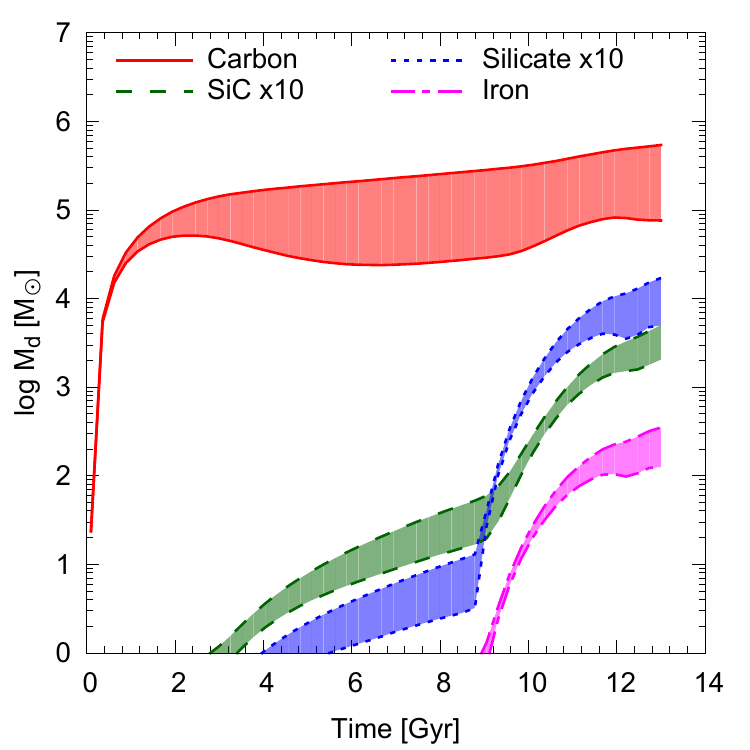}
\caption{\textit{Left panel:} Time evolution of dust production rates by AGB stars for silicate, carbonaceous, SiC and iron dust and by type II SNe for silicates and carbonaceous dust in the LMC. Open and closed symbols show the current DPRs for carbon- and oxygen-rich dust, respectively, derived from IR observations (\cite{Matsuura:2009p363} - triangles, with an arrow marking an upper limit, \cite{Boyer:2012ck} - squares, \cite{Riebel:2012eq} - circles and \cite{Srinivasan:6p7509} - diamonds). \textit{Right panel:}  Time evolution of accumulated dust mass of dust species ejected by AGB stars over the LMC history. Lower and upper borders of shaded areas denote the calculations  with and without destruction processes. The data for plots are taken from \cite{Zhukovska:2013vg}.}
\label{fig:dustinjrates}
\end{figure}

\subsubsection{Magellanic Clouds}
Findings from the dust evolution models that stars contribute a minor fraction to the ISM dust reservoir in our galaxy cannot be checked directly with observations because of the large extinction in the galactic disk. The Magellanic Clouds are the first galaxies for which dust mass-loss rates have been measured from IR observations for the entire population of evolved stars.  The Clouds thus offer a chance to compare the synthesised present population of AGB stars, and the net production rates and composition of the ejected dust with observations.  This became possible with the \textit{Spitzer} Legacy Programs ``Surveying the Agents of Galaxy Evolution in the LMC'' (SAGE-LMC) that catalogued over 8 million IR sources \cite{2006AJ....132.2268M} and the SAGE-SMC program that catalogues  over 2 million IR sources in the SMC \cite{Gordon:2011jq}. 

The SAGE survey quantifies the dust mass-loss rates by detecting excess emission at 8 and 24~\mum. Several methods and source classifications  have been used to measure individual dust production rates (DPR) by evolved stars from SAGE-LMC data \cite{Matsuura:2009p363, Boyer:2012ck, Riebel:2012eq, Srinivasan:6p7509}. Their measurements for carbon-rich and oxygen-rich AGB stars in the Magellanic Clouds together with theoretical predictions are given in Table~\ref{tab:MCs_DPR}.
One way to measure the dust mass-loss rates from AGB stars is by means of empirical relations between dust-mass loss rate and excess dust emission in the mid-IR  \cite{Srinivasan:6p7509, Matsuura:2009p363, Boyer:2012ck}. The IR excess from circumstellar shells is calculated by comparing the total flux from the source observed in each band to the flux expected from the central star. The DPR for the evolved stellar population in the LMC derived from the IR excess is dominated by a small number of highly evolved stars undergoing heavy mass-loss (extreme sources). Extreme AGB stars produce dust at the rate of $0.9-2.4\times10^{-5}\Msyr$, this is about ten times faster than the DPR by low- and moderately-obscured AGB stars \cite{Srinivasan:6p7509, Boyer:2012ck}. It is not possible to  separate extreme sources into stars with O- and C-rich chemistries based on photometry only because of their heavy obscuration, but the spectroscopic analysis for their subsample indicates that most of them are C-rich \cite{Srinivasan:6p7509}. \cite{Matsuura:2009p363} adopted a different empirical relation between IR excess and mass-loss rate and an alternative source classification separating C-rich from O-rich stars, which led them to a higher DPR of $\gtrsim4\times10^{-5}\Msyr$ for carbon-stars and a lower limit of $4\times10^{-7}\Msyr$ for oxygen-rich stars.
A more elaborate method to estimate the DPR of an AGB star involves fitting its spectral energy distribution with a pre-computed grid of radiative transfer models of evolved stars. Such Grid of Red Supergiant and Asymptotic Giant Branch ModelS (GRAMS)  designed for C-  and O-rich stars was applied to fit the multi-band photometry of $\sim30000$ AGB and RSG stars in the LMC and measure their individual DPRs and bolometric luminosities \cite{Riebel:2012eq}. This work managed to discriminate between O- and C-rich extreme stars based on the chemistry type of the best fit GRAM model. They confirm that majority of extreme stars ( 97\%) are indeed C-rich and they dominate the dust input from evolved stars (75\% of total DPR). The models of the AGB dust populations in the LMC indicate that extreme stars are likely AGB stars in the super-wind evolutionary phase.

Dust input from AGB stars during the LMC history has been recently investigated with a model of dust evolution and mass- and metallicity-dependent dust yields \cite{Zhukovska:2013vg}. The model dust mixture is dominated by carbonaceous grains ejected at the rate $0.2-1\times 10^{-4}\Ms$ for the whole LMC evolution (Fig.~\ref{fig:dustinjrates}). The production of silicate, iron and SiC dust is very inefficient until 4~Gyr ago, when it rapidly increased to its present level of $10^{-6}\Msyr$. This rise is related to the dust return from the generation of AGB stars formed out of the ISM which was enriched by recent bursts of star formation. The silicate-to-carbon dust mass ratio in the stardust population from AGB stars in the LMC survived until the present day is also plotted for $Z_{\rm LMC}=0.008$ in Fig.~\ref{fig:AGBradial}. Despite the different star formation histories of the LMC and the Milky Way, the ratio in the LMC is only slightly higher than that at the Galactic radius with $Z=Z_{\rm LMC}$.
The injection rates for carbonaceous and silicate dust are commensurate with the values for C- and O-rich dust derived from the IR observations, within the scatter of various observational estimates (Fig.~\ref{fig:dustinjrates}). Comparison of the parent stellar population of silicate and iron grains from AGB stars in the LMC and the local Milky Way reveals drastic differences stemming from the strong metallicity dependence. Most of silicate and iron grains originate from a small population of intermediate-mass stars consisting of only 4\% of the total number of stars, whereas in the solar neighbourhood they originate from low-mass stars.

\section{Can stellar sources reproduce the interstellar dust budget?}\label{sec:Final}
The models of the lifecycle of dust predict that AGB stars in the solar neighbourhood contribute only a minor fraction of roughly 1.5\% for carbonaceous and $\lesssim 1\%$ for silicate dust, respectively, to the interstellar dust budget \cite{Zhukovska:2008bw}. The contribution from SNe is also very small, of about 1\%. Although the stardust fraction increases towards the outer disk, dust grown by accretion in the dense ISM dominates the interstellar dust population in the Milky Way. The stardust constitutes only a small fraction of the total dust mass \cite{Zhukovska:2008p7215, Zhukovska:2009p7232}.

How does the contribution from stars change at subsolar metallicities? 
We can answer this question by comparing the dust mass accumulated over the LMC history predicted by the models, which have been verified with observations at the present time, and compare it with the existing dust mass in the ISM of $(1.1-2.5) \times 10^6 \Ms$ as estimated from extinction and emission studies.  Figure~\ref{fig:dustinjrates} shows time evolution of the total stardust masses for the main dust species from AGB stars in the LMC  calculated for two limiting cases: with and without dust destruction in the ISM. The upper limit of the current dust mass form AGB stars of $5\times 10^5\Ms$ is lower than the present interstellar dust mass. After 0.1-1~Gyr of evolution, the dust-to-gas ratio in the LMC is dominated by the ISM-grown dust, with AGB stars and SNe contributing about 10\% (Fig.~\ref{fig:DGR}). This result confirms a large discrepancy between dust input from stars and the existing interstellar dust mass in the LMC termed as ``missing dust-mass problem'' in \cite{Matsuura:2009p363}. Agreement between the models of the lifecycle of stardust grains in the LMC and observations on dust production by stellar sources in the dust mass budget implies that dust growth by accretion in the ISM, a major dust source in the solar neighbourhood, is also important at subsolar metallicities of the LMC \cite{Zhukovska:2013vg}. 

The dust mass budget at the lower metallicity of the SMC has been studied observationally within the SAGE-SMC programs \cite{Boyer:2012ck, 2013MNRAS.429.2527M}. AGB stars contribute only about 2\%  to the interstellar dust mass in the SMC of $(0.29-1.1) \times 10^6\Ms$, estimated from IR and submillimeter imaging \cite{Boyer:2012ck} and references therein). The star formation rate has increased by a factor of 4 during the last 12~Myr resulting in the higher relative SN rate in the SMC  compared to the Milky Way and the LMC and, consequently, enhanced dust production by SNe. Considering upper and lower limits for SN dust production and destruction,  \cite{Boyer:2012ck} find that the interstellar dust can be accounted for solely by stellar sources if all SNe produce dust mass of at least $0.05\Ms$ and if most SNe explode in dense regions where much of the ISM dust is shielded from the shocks. This scenario is however not likely, since 20-25\% of massive stars are ejected from their parent clusters and become runaway stars \cite{Oey:2012ts}. Travelling often at supersonic velocities, these stars  end their life far away from their natal dense clouds and explode as isolated SNe in the warm or hot gas, which have much larger filling factors than molecular clouds. The question on whether destruction by the isolated SNe can balance dust input from type II SNe in the SMC can be answered in future with dust evolution models. Generic models of dust evolution in late-type dwarf galaxies indicate that, despite the longer destruction timescales in gas-rich dwarf galaxies,  the dust growth by accretion in the ISM does plays an important role even in metal-poor galaxies \cite{Zhukovska:2013vg}. The characteristic timescales on which this process becomes important in these galaxies are $\lesssim 1$~Gyr, which is longer than in the local Milky Way ( $\lesssim 1$~Myr), but significantly shorter than the age of old stellar populations in the SMC implying that the dust growth should be important in the SMC at present epoch.

\paragraph*{Acknowledgements.} S.Z. thanks the organisers of The Life Cycle of Dust in the Universe meeting in Taipei for their hospitality and stimulating and productive atmosphere during the meeting. S.Z. acknowledges support by	 the Deutsche Forschungsgemeinschaft through SPP 1573: ``Physics of the Interstellar Medium''. 


\end{document}

%% file: journals.tex
\def\aj{AJ}%
\def\actaa{Acta Astron.}%
\def\araa{ARA\&A}%
\def\apj{ApJ}%
\def\apjl{ApJ}%
\def\apjs{ApJS}%
\def\ao{Appl.~Opt.}%
\def\apss{Ap\&SS}%
\def\aap{A\&A}%
\def\aapr{A\&A~Rev.}%
\def\aaps{A\&AS}%
\def\azh{AZh}%
\def\baas{BAAS}%
\def\bac{Bull. astr. Inst. Czechosl.}%
\def\caa{Chinese Astron. Astrophys.}%
\def\cjaa{Chinese J. Astron. Astrophys.}%
\def\icarus{Icarus}%
\def\jcap{J. Cosmology Astropart. Phys.}%
\def\jrasc{JRASC}%
\def\mnras{MNRAS}%
\def\memras{MmRAS}%
\def\na{New A}%
\def\nar{New A Rev.}%
\def\pasa{PASA}%
\def\pra{Phys.~Rev.~A}%
\def\prb{Phys.~Rev.~B}%
\def\prc{Phys.~Rev.~C}%
\def\prd{Phys.~Rev.~D}%
\def\pre{Phys.~Rev.~E}%
\def\prl{Phys.~Rev.~Lett.}%
\def\pasp{PASP}%
\def\pasj{PASJ}%
\def\qjras{QJRAS}%
\def\rmxaa{Rev. Mexicana Astron. Astrofis.}%
\def\skytel{S\&T}%
\def\solphys{Sol.~Phys.}%
\def\sovast{Soviet~Ast.}%
\def\ssr{Space~Sci.~Rev.}%
\def\zap{ZAp}%
\def\nat{Nature}%
\def\iaucirc{IAU~Circ.}%
\def\aplett{Astrophys.~Lett.}%
\def\apspr{Astrophys.~Space~Phys.~Res.}%
\def\bain{Bull.~Astron.~Inst.~Netherlands}%
\def\fcp{Fund.~Cosmic~Phys.}%
\def\gca{Geochim.~Cosmochim.~Acta}%
\def\grl{Geophys.~Res.~Lett.}%
\def\jcp{J.~Chem.~Phys.}%
\def\jgr{J.~Geophys.~Res.}%
\def\jqsrt{J.~Quant.~Spec.~Radiat.~Transf.}%
\def\memsai{Mem.~Soc.~Astron.~Italiana}%
\def\nphysa{Nucl.~Phys.~A}%
\def\physrep{Phys.~Rep.}%
\def\physscr{Phys.~Scr}%
\def\planss{Planet.~Space~Sci.}%
\def\procspie{Proc.~SPIE}%
\let\astap=\aap
\let\apjlett=\apjl
\let\apjsupp=\apjs
\let\applopt=\ao